# Selective enhancement of resonant multiphoton ionization with strong laser fields

Min Li,[1] Peng Zhang,[1] Siqiang Luo,[1] Yueming Zhou,[1] Qingbin Zhang,[1] Pengfei Lan,[1,*] and Peixiang Lu[1,2,†]

[1]*School of Physics and Wuhan National Laboratory for Optoelectronics, Huazhong University of Science and Technology, Wuhan 430074, People's Republic of China*
[2]*Laboratory of Optical Information Technology, Wuhan Institute of Technology, Wuhan 430073, People's Republic of China*

High-resolution photoelectron momentum distributions of Xe atom ionized by 800-nm linearly polarized laser fields have been traced at intensities from $1.1\times10^{13}$ W/cm$^2$ to $3.5\times10^{13}$ W/cm$^2$ using velocity-map imaging techniques. At certain laser intensities, the momentum spectrum exhibits a distinct double-ring structure for low-order above-threshold ionization, which appears to be absent at lower or higher laser intensities. By investigating intensity-resolved photoelectron energy spectrum, we find that this double-ring structure originates from resonant multiphoton ionization involving multiple Rydberg states of atoms. Varying the laser intensity, we can selectively enhance the resonant multiphoton ionization through certain atomic Rydberg states. The photoelectron angular distributions of multiphoton resonance are also investigated for the low-order above threshold ionization.

PACS number(s)：32.80. Fb, 32.80.Rm

---

* pengfeilan@mail.hust.edu.cn
† lupeixiang@mail.hust.edu.cn

When an atom is exposed to a strong laser field, an electron can be ionized with absorption of multiple photons beyond the ionization threshold. This process, known as above-threshold ionization (ATI), was initially observed more than three decades ago [1] (for a review, see [2]). The observation of the ATI has been regarded as a benchmark experiment in strong-field physics. As one of the most fundamental strong-field processes, ATI has attracted widespread interest over the past years. The ATI process is governed by much the same physics as some other interesting strong-field phenomena, such as high-order harmonic generation and double ionization.

Generally, ATI can be interpreted as multiphoton absorption via nonresonant or resonant states when the Keldysh parameter [3] $\gamma$ is larger than one [$\gamma = \sqrt{I_p/2U_p}$ where $I_p$ is the ionization potential, $U_p = F_0^2/4\omega^2$ the ponderomotive energy, $F_0$ the field amplitude, $\omega$ the laser frequency]. Atomic units (a.u.) are used throughout unless stated otherwise. For the nonresonant ionization, the formation of the photoelectron momentum distributions and angular distributions in ATI has been revealed in terms of the interferences of electron trajectories [4-7]. Because of the ponderomotive shift of Rydberg state close to the ionization limit, the atom can be resonantly excited to the Rydberg state with absorption of several photons and subsequently be ionized by one or more additional photons. This resonant multiphoton ionization process, dubbed Freeman resonance [8,9], has a much larger ionization probability as compared with the nonresonant ionization channel. The Freeman resonance manifests itself as many discrete sub-structures from different Rydberg states for the low order ATI peaks in the photoelectron energy spectrum when the pulse duration is short [8]. The photoelectron angular distributions (PADs) of the Freeman resonance are believed to have relations with the angular momentum quantum number of the intermediate state [10-13].

Both resonant ionization and nonresonant ionization can significantly contribute to the final electron momentum spectrum and their contributions will mix each other in the momentum spectrum. Recently, Shao *et al*. [14] have isolated the resonantly-enhanced multiphoton ionization from the nonresonant contributions using intensity-resolved photoelectron energy spectrum. The isolation is based on the fact that the nonresonant ATI peak shifts towards lower energy with the increase of the laser intensity while the resonant ionization is independent of the laser intensity [15]. It is shown that the intensity-resolved energy and momentum spectra offer a wealth of information about the formation of the ATI. Actually, the intensity-resolved energy or momentum spectrum has been used to investigate the underlying dynamics of single ionization [14,16-18], double ionization [19], and high-order harmonic generation [20]. For single ionization, the intensity-resolved energy spectrum has only studied at comparably large intensities (larger than $3\times10^{13}$ W/cm$^2$ for Xe and larger than $1\times10^{14}$ W/cm$^2$ for H) [14,16,17], at which the nonresonant ionization has a significant contribution. Decreasing the laser intensity, the relative contribution of the nonresonant multiphoton ionization decreases [21]. Moreover, at a low laser intensity, the focal volume effect of the laser beam plays a minor role in the experiment. Only one or a few Rydberg states might come into resonance. Therefore, it is possible to selectively enhance the resonant multiphoton ionization via certain Rydberg states from the contributions of the nonresonant ionization by tuning the laser intensity.

In this work, we measure high-resolution two-dimensional photoelectron momentum distributions from single ionization of xenon atoms in a linearly polarized laser pulse at intensities of 1.1-3.5×10$^{13}$W/cm$^2$ using a velocity-map imaging (VMI) spectrometer. VMI is widely used in atomic, molecular and optical science. Compared with other momentum imaging techniques, e.g., cold target recoil ion momentum spectroscopy (COLTRIMS) setup [22], VMI has a significantly higher count rate per pulse and thus the signal to noise ratio of the VMI is better than that of the COLTRIMS setup. Benefiting from this advantage, we have precisely measured the two-dimensional photoelectron momentum distributions to obtain deep insight into the multiphoton ionization of Xe at low laser intensities. We find that the contribution of the resonant ionization is dominant over that of the nonresonant ionization at low laser intensities. More interestingly, the recorded velocity map shows a distinct double-ring structure at certain laser intensities. The inner ring structure disappears at lower laser intensities while the outer ring structure disappears at higher laser intensities. By studying the intensity-resolved photoelectron energy spectrum, we find that this double-ring structure results from resonant multiphoton excitation of two Rydberg $f$ states of Xe atom. Varying the laser intensity, we can selectively control different atomic Rydberg states via which the resonantly-enhanced multiphoton ionization occurs. We have also studied the PADs of multiphoton resonances for low-order above threshold ionization. We find that the PADs of multiphoton resonance depend not only on the angular momentum quantum number but also on the principal quantum number of the Rydberg states.

Experimentally, we use ~35 fs FWHM laser pulses centered at 800 nm from a Ti: sapphire laser system at a repetition of 1 kHz to ionize xenon atoms. The amplified pulse energy can be up to 10 mJ. The pulses are focused into the extraction region of a vacuum chamber of a homebuilt VMI spectrometer [23,24] using a convex lens ($f$ = 40 cm). Electrons that are ionized at the crossing point of the laser beam and an atomic beam are subsequently focused by a static electric field onto a dual multichannel plate (MCP) detector, followed by a phosphor screen and a CCD camera system. A voltage of 1.3 kV is applied over the MCPs. At the rear side of the MCP, the electrons are further accelerated up to 4.7 kV towards the phosphor screen. To obtain high-resolution spectra for the two-dimensional photoelectron momentum distributions of the low-order ATI, we put a voltage of -500 V on the repeller electrode and -390 V on the extractor electrode to collect the photoelectrons, corresponding to a small electric field of ~83 V/cm at the laser focus position. The impact of the electrons on the phosphor screen produces the fluorescence, which is recorded by the computer-controlled CCD camera. The measured raw photoelectron images are inverted using the Gaussian basis-set expansion Abel transform method [25]. Then the inverted images are calibrated by means of the ATI peak separation that is equal to the photon energy.

The background pressure of the vacuum chamber is about $4\times10^{-8}$ mbar. To reduce the contribution of the electrons generated from the background gas, both the images with and without xenon in the chamber are measured under the same experimental conditions. The final spectrum is obtained by subtracting the image without xenon from the corresponding image measured with xenon. The polarization of the laser beam is parallel to the MCP detector and perpendicular to the static electric field. The laser intensity is changed by rotating a $\lambda/2$ wave plate before a polarizer. We calibrate the laser intensity by the energy shift of the nonresonant ATI peaks, which is proportional to $U_p$ [14]. The uncertainty of the intensity calibration is estimated to be ~5%.

Figures 1(a)-1(c) show the measured two-dimensional photoelectron momentum distributions of Xe atoms in a strong laser field at intensities of $1.3\times10^{13}$W/cm$^2$, $1.6\times10^{13}$W/cm$^2$, and $1.9\times10^{13}$W/cm$^2$, respectively. The noise in the central line of $p_z=0$ is due to the numerical noise generated in the Abel transformation algorithm [25]. One can find that the photoelectron momentum spectra show many ring-like structures centered around zero momentum, which correspond to the ATI peaks in the energy spectra, as seen in Figs. 1(d)-1(f). Within each order ATI ring structure, there are many spot-like structures. For the resonant multiphoton ionization, these spot-like structures are related to the angular momentum quantum number of the intermediate Rydberg state while for the nonresonant ionization, these spot-like structures can be interpreted as a result of interference among electron wave packets within each laser cycle [7].

Interestingly, at the intensity of $1.6\times10^{13}$W/cm$^2$ the momentum spectrum exhibits a pronounced double-ring structure, as seen in Fig. 1(b). The angular distributions of these two rings look quite similar. The resulting photoelectron energy distribution in Fig. 1(e) also reveals a double-peak structure within each order ATI, e.g., the peaks of ~0.75eV and ~1.05eV for the first order ATI. These two peaks correspond to the inner ring and the outer ring in Fig. 1(b), respectively. At the lower laser intensity of $1.3\times10^{13}$W/cm$^2$, only the outer ring structure (~1.05 eV in the energy spectrum) appears while the inner ring structure (~0.75 eV in the energy spectrum) is absent. The case is reversed for the higher laser intensity of $1.9\times10^{13}$W/cm$^2$, where the inner ring structure can be clearly seen and the outer ring structure nearly disappears. Similar double-ring structures can be found in recent experiments [26,27], but little attention has been paid to the underlying mechanism.

To shed light on the underlying mechanism of the double-ring structure, we have recorded the velocity map of photoelectron from single ionization of xenon with the laser intensities changing from $1.1\times10^{13}$W/cm$^2$ to $3.5\times10^{13}$W/cm$^2$. The intensity scan corresponds to the range of the Keldysh parameter $1.7<\gamma<3.0$. We show in Fig. 2 the resulting photoelectron angle-integrated energy spectra from $1.3\times10^{13}$W/cm$^2$ to $2.0\times10^{13}$W/cm$^2$. One can clearly see that the positions of these two peaks of ~0.75 eV and ~1.05 eV do not shift with the change of the laser intensity (guided by the vertical dashed lines), which indicates that they come from the Freeman resonance. To find the intermediate Rydberg states of these two peaks, we refer to the energy level of Xe atom from the National Institute of Standards and Technology (NIST) [28]. We find that one-photon ionization from the 4$f$ and 5$f$ states will contribute to the prominent electron peaks at ~0.75 eV (inner ring) and ~1.05 eV (outer ring), respectively. At a low laser intensity of $1.3\times10^{13}$W/cm$^2$, the energy spectrum is dominated by the resonant ionization via the 5$f$ state (~1.05 eV). With increasing the laser intensity, the relative contribution of the 5$f$ resonant ionization deceases and the relative contribution of the 4$f$ resonant ionization (~0.75 eV) increases. At a moderate intensity of $1.6\times10^{13}$W/cm$^2$, both the 5$f$ and 4$f$ Rydberg states come into resonance and these two ionization channels have a comparable contribution to the final momentum distributions. Due to the same angular momentum quantum number, the angular distribution of the multiphoton ionization from these two resonant states looks similar, thus leading to the double-ring structure in the photoelectron momentum spectrum.

In Fig. 3, we show the intensity-resolved photoelectron energy spectrum spanning a larger range of laser intensities from $1.1\times10^{13}$W/cm$^2$ to $3.5\times10^{13}$W/cm$^2$. The yield (color scale) is normalized to the maximum value at each laser intensity. One can find that the electron energy peaks shift towards lower energy as the intensity increased. This intensity dependence agrees well with the previous experiment and simulation [14,16]. The reason is that the ionization threshold has an ac-Stark shift that is equal to the ponderomotive energy. Thus the effective ionization potential increases by $U_p$ and the expected ATI peaks for the nonresonant ionization in the photoelectron spectrum satisfy,

$$E_k = n\omega - (I_p + U_p) \tag{1}$$

where $n$ is the total number of photons absorbed in the ionization process. The dashed lines in Fig. 3 show the expected ATI peaks of the nonresonant ionization for $n$-photon ionization ($n$=9, 10, 11) according to Eq. (1). We have used this intensity-dependent feature to calibrate the peak laser intensity [14]. In addition, in Fig. 3, one can see that there are many other pronounced structures embedded in the contributions of the nonresonant ionization, which are independent on the laser intensity. These structures are related to the resonantly enhanced multiphoton ionization via the Rydberg states. Compared with the result in Ref. [14], the contribution of the resonant ionization is dominant over that of the nonresonant ionization at low laser intensities.

By comparing the ATI peak position with the available database of Xe atom [28], we also show the energy levels of the intermediate states that correspond to those intensity-independent enhanced-ionization peaks in Fig. 3. From Fig. 3, one can see that the ionization via the 6$f$ state is resonantly enhanced at a low laser intensity of ~$1.1 \times 10^{13}$ W/cm$^2$. When the laser intensity increases to ~$1.15 \times 10^{13}$ W/cm$^2$, the contribution of the resonant ionization via the 6$f$ state is highly suppressed and the relative contribution of the 5$f$ resonant ionization increases. Increasing the intensity to ~$1.5 \times 10^{13}$ W/cm$^2$, the 4$f$ state comes into resonance and the relative contribution of the 5$f$ resonant ionization begins to decrease. When the intensity becomes ~$2.6 \times 10^{13}$ W/cm$^2$, the resonant ionization through the 7$p$ state also contributes significantly to the energy spectrum. Further increasing the laser intensity (~$3 \times 10^{13}$ W/cm$^2$), the nine-photon ionization channel closes. After the intensity of ~$3.2 \times 10^{13}$ W/cm$^2$, the 5$g$ and 6$g$ states are shifted into resonance via a nine-photon absorption from the ground state. Therefore, through tuning the laser intensity, we can select certain Rydberg states through which the resonantly-enhanced multiphoton ionization occurs.

The selective enhancement can be interpreted within the multiphoton absorption picture. For the resonant ionization, an integer number of photons are absorbed to excite the electron from the ground state to a Rydberg state. Due to the ponderomotive shift of the weakly bound Rydberg state, only a certain intensity component will satisfy the resonant condition for a specific Rydberg state, i.e.,

$$m\omega = \frac{I_r}{4\omega^2} + I_p - E_r \qquad (2)$$

where $E_r$ is the energy of the Rydberg state, $m$ is the number of photons absorbed from the ground state to the Rydberg state, and $I_r$ is the corresponding resonant laser intensity. Because of the dipolar parity selection rules [26], the excitation of the *p* and *f* states need to absorb an even number of photons and the excitation of the *s*, *d* and *g* states need to absorb an odd number of photons for xenon atom. Table I shows the resonant laser intensities for different Rydberg states of Xe via an eight-photon ($m=8$) or nine-photon ($m=9$) excitation calculated according to Eq. (2) for an 800-nm, 35-fs laser pulse. The predicted resonant laser intensity at 800 nm agrees well with the measurement in Fig. 3. For example, as seen in Fig. 3, the laser intensities for the resonantly enhanced multiphoton ionization through the 5*f* and 4*f* states are about $(1.2\text{-}1.65)\times10^{13}$W/cm$^2$ and $(1.45\text{-}3.4)\times10^{13}$W/cm$^2$, in a good agreement with the prediction of $1.38\times10^{13}$W/cm$^2$ and $1.91\times10^{13}$W/cm$^2$, respectively. Due to the focal volume effect of the laser beam in the experiment, the resonant multiphoton ionization via an intermediate state occurs at a certain range of laser intensities, not only a single laser intensity.

TABLE I. Resonant laser intensity $I_r$ for different Rydberg states calculated according to Eq. (2) for an 800-nm laser pulse

| 8-photon excitation | 6*f* | 5*f* | 4*f* | 7*p* |
|---|---|---|---|---|
| $I_r$ (W/cm$^2$) | $1.09\times10^{13}$ | $1.38\times10^{13}$ | $1.91\times10^{13}$ | $2.52\times10^{13}$ |
| 9-photon excitation | 6*g* | 5*g* | | |
| $I_r$ (W/cm$^2$) | $3.69\times10^{13}$ | $3.97\times10^{13}$ | | |

Because the PADs can offer more information about the ATI process [26], we further show in Fig. 4(a) the PADs of the first-order ATI peak of Fig. 1(a) and Fig. 1(c), corresponding to the intensities of $1.3\times10^{13}$W/cm$^2$ and $1.9\times10^{13}$W/cm$^2$, respectively. The PADs show the main lobes in the laser polarization direction (0° and 180°) and several pronounced jet structures (~40°, ~90°, and ~140°) sticking out from the waist of the main lobes [29]. Theoretically, the PADs with an angular momentum number of $l$ can be expressed as [4],

$$\frac{d^2P}{dEd(\cos\theta)} \approx [P_l(\cos\theta)]^2 \qquad (3)$$

where $P_l(\cos\theta)$ is a Legendre polynomial of degree $l$ and $E$ is the electron energy. We show the distribution of $[P_4(\cos\theta)]^2$ in Fig. 4(a) by the blue curve. One can see that the measured PADs of the first-order ATI show three-jet structures in each side of the PADs, the same number as that of $[P_4(\cos\theta)]^2$. This reveals that the dominant angular momentum of those photoelectrons is four and those photoelectrons come from one-photon ionization of an intermediate $f$ state. In the above analysis, we have shown that they indeed come from the resonantly-enhanced ionization via the intermediate 5$f$ and 4$f$ states at the intensities of $1.3\times10^{13}$W/cm$^2$ and $1.9\times10^{13}$W/cm$^2$, respectively. Close inspection of the PADs reveals that there is an evident difference of the jet height between the 5$f$ and 4$f$ resonant ionization, though their dominant angular momenta are the same. The jet structure at ~90° is more pronounced for the 5$f$ resonant ionization while the jet structures at ~40° and ~140° are more pronounced for the 4$f$ resonant ionization.

To see how the PADs change with the laser intensity, we show the PADs for the first-order ATI in Fig. 4(b) for the intensities from $1.3\times10^{13}$W/cm$^2$ to $2.0\times10^{13}$W/cm$^2$. The laser intensity of each curve in Fig. 4(b) is the same as that in Fig. 2. The positions of these jet structures do not shift with the laser intensity. However, the relative height of the jet structure is changed as the laser intensity increases. The jet structure at 90° is more obvious at lower laser intensities, whereas the jet structures at ~40° and ~140° are more obvious at higher laser intensities. An evident change of the jet heights occurs at the intensity of ~$1.6\times10^{13}$W/cm$^2$, which corresponds to the transition from the 5*f* resonant ionization to the 4*f* resonant ionization (see Fig. 2). The difference between the PADs of the 5*f* and 4*f* resonant ionization indicates that the principal quantum number of the Rydberg state plays an important role in the formation of the PADs of the Freeman resonance.

Note that the single Legendre polynomial is just a single partial wave for the angular distribution. The PADs in Fig. 4 is dominated by the angular momentum of four, but other partial waves may also contribute to the PADs. The angular distribution of the 5*f* resonant ionization looks also like the distribution of $[P_2(\cos\theta)]^2$, which means that the partial wave of the angular momentum of two has a significant contribution to the angular distribution. In fact, for an intermediate state with an angular momentum of *l*, the angular momentum of one-photon ionization from this state can be *l*+1 or *l*-1. Generally, the transition matrix from *l* to *l*+1 is larger than the transition matrix from *l* to *l*-1 [30]. Therefore the PADs of the one-photon ionization from an *f* state is dominated by the contribution of *l*=4. The transition matrix from *l*=3 to *l*=2 for the 5*f* resonant ionization might be larger than that of 4*f* resonant ionization. Thus the relative contribution of *l*=2 is larger for the 5*f* resonant ionization, as compared with the 4*f* resonant ionization.

Fig. 5(a) and Fig. 5(b) show the PADs from the (8+2)-photon ionization via the resonant 4*f* and 5*f* states and from the (9+1)-photon ionization via the resonant 5*g* and 6*g* states, respectively. As seen in Fig. 5(b), the PADs of the (9+1)-photon ionization via the 5*g* and 6*g* states reveal four jets in each side of the plane, the same number as that of $[P_5(\cos\theta)]^2$. Thus, their dominant angular momenta are both five. For the PADs of the (8+2)-photon ionization via the 4*f* and 5*f* ionization [Fig. 5(a)], only two jets can be clearly seen in each side of the plane. However, the positions of these jet structures agree well with $[P_5(\cos\theta)]^2$ instead of $[P_3(\cos\theta)]^2$, as seen in Fig. 5(a). Therefore, those electrons also have a dominant angular momentum of five. Because those electrons come from the two-photon ionization from the 4*f* and 5*f* states, the partial waves of *l*=1 and *l*=3 also have a significant contribution, which might reduce the height of the jet structure at ~40° as compared with the distribution of $[P_5(\cos\theta)]^2$. From Fig. 5, one can still see that the position of the jet structure is nearly the same for the resonant ionization via different intermediate states when the dominant angular momentum is the same. However, the height of the jet structure is different between the 4*f* and the 5*f* resonant ionization or between the 5*g* and the 6*g* resonant ionization. The differences become smaller as compared with the case of (8+1)-photon ionization in Fig. 4(a).

In conclusion, we have systematically studied two-dimensional photoelectron momentum distributions of xenon atom in strong laser fields at intensities of $1.1-3.5\times10^{13}$W/cm$^2$ using the velocity-map imaging spectrometer. The recorded images show a pronounced double-ring structure for low-order above-threshold ionization at certain laser intensities. The inner ring disappears at lower laser intensities while the outer ring disappears at higher laser intensities. By investigating intensity-resolved photoelectron energy spectrum, we find that this double-ring structure is a result of resonant multiphoton ionization involving two Rydberg states of atoms. Varying the laser intensity, we can selectively enhance multiphoton resonant ionization via specific atomic Rydberg states compared with the contribution of the nonresonant ionization. We have also studied the intensity-dependent photoelectron angular distributions of multiphoton resonances for low-order above threshold ionization. We find that the photoelectron angular distribution of multiphoton resonance depends not only on the angular momentum quantum number but also on the principal quantum number of the Rydberg states. This study enriches the understanding of strong-field multiphoton ionization and might stimulate experimental and theoretical interests in coherent control of electron dynamics using strong laser fields.

This work was supported by the National Natural Science Foundation of China under Grants No. 61405064, No. 11234004 and No. 11422435.

**FIGURE CAPTION**:

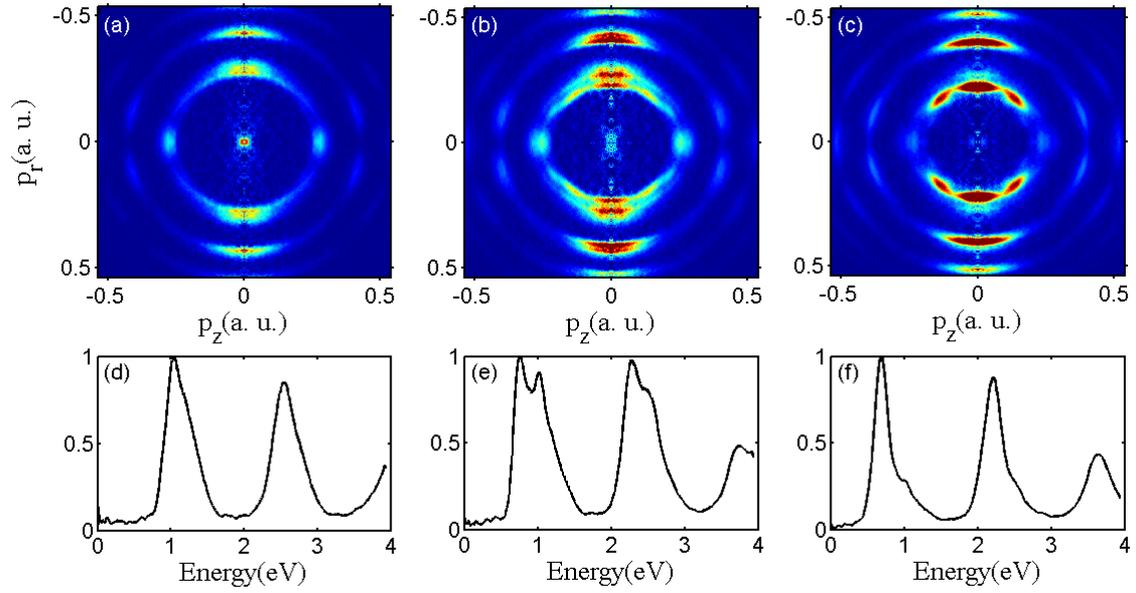

FIG. 1. (Color online) (a)-(c) show the two-dimensional photoelectron momentum distributions of xenon atom in an 800-nm laser field at intensities of (a) $1.3\times10^{13}$W/cm$^2$, (b) $1.6\times10^{13}$W/cm$^2$, and (c) $1.9\times10^{13}$W/cm$^2$, respectively. The laser polarization direction is along the vertical axis. (d)-(f) show the corresponding angle-integrated energy spectra of (a)-(c).

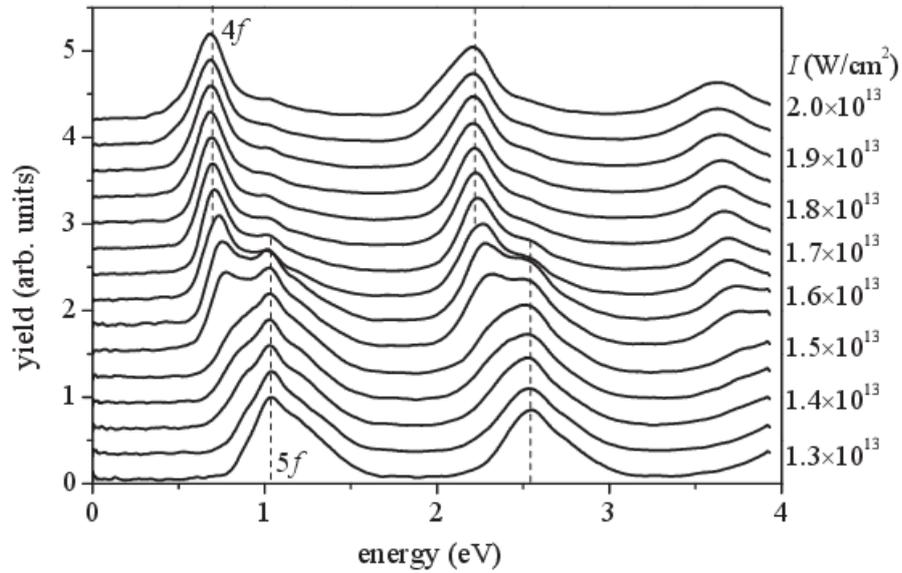

FIG. 2. The angle-integrated energy spectra of the first two order ATI of Xe for intensities ranging from $1.3\times10^{13}$W/cm$^2$ to $2.0\times10^{13}$W/cm$^2$. Vertical dashed lines are used to guide the first and second order ATI peaks.

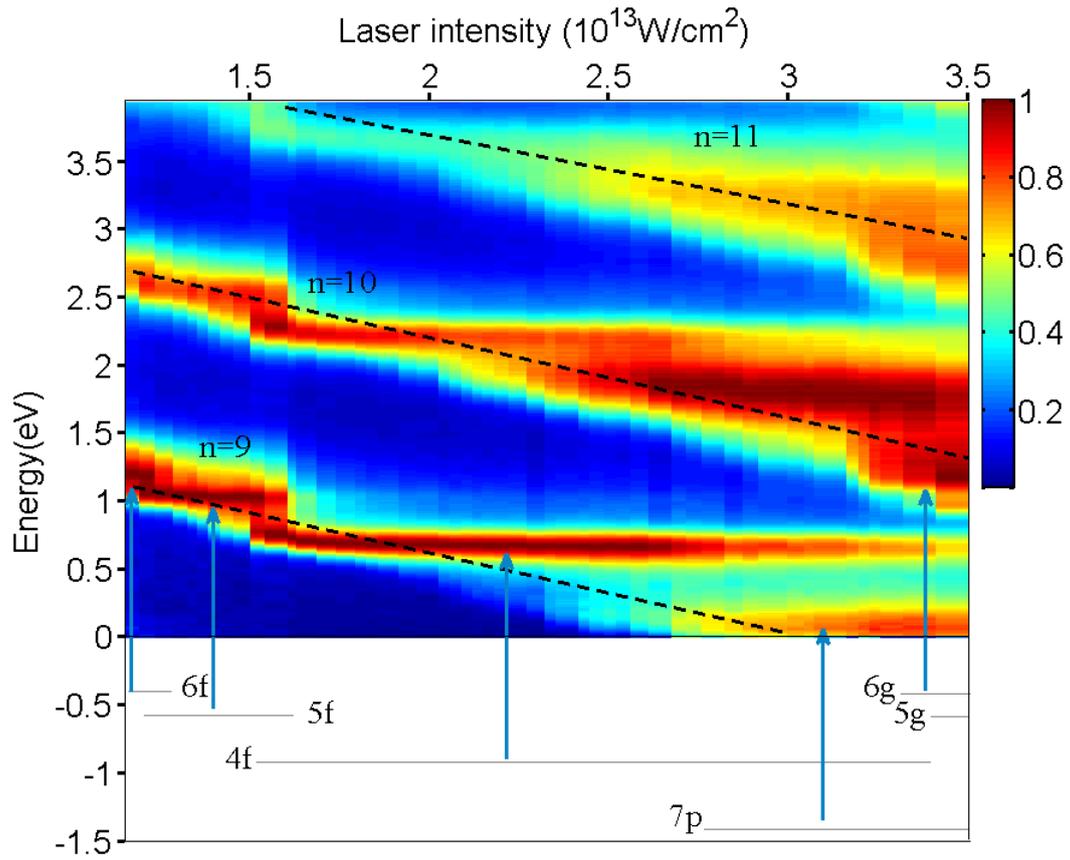

FIG. 3. (Color online) The intensity-resolved photoelectron energy distributions for the intensities from $1.1\times10^{13}$W/cm$^2$ to $3.5\times10^{13}$W/cm$^2$. The color scale is normalized to the maximum yield at each laser intensity. The dashed lines denote the $n$-photon ATI peaks according to Eq. (1). The solid horizontal lines with negative energy show the energy level of the resonant states involved in the experiment.

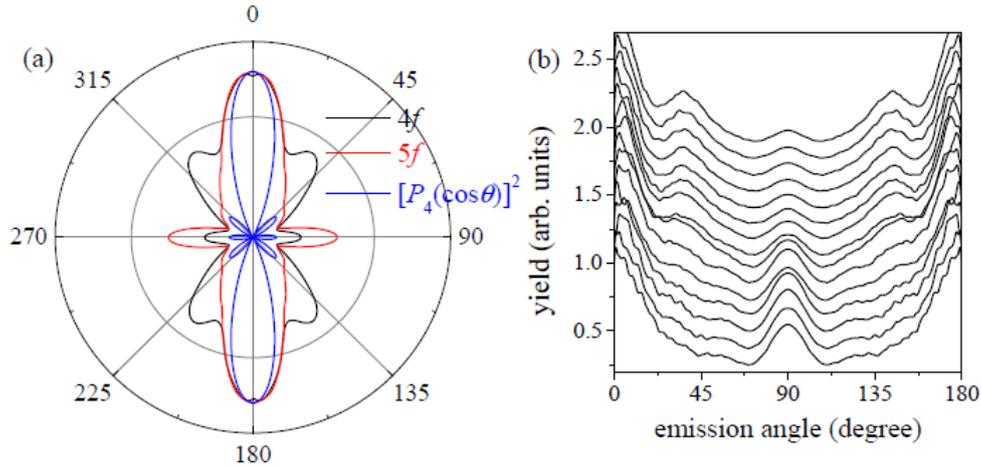

FIG. 4. (Color online) (a) Polar plot of the photoelectron angular distributions of the first-order ATI peak at intensities of $1.3\times10^{13}$W/cm$^2$ and $1.9\times10^{13}$W/cm$^2$ extracted from Figs. 1(a) and 1(c), which correspond to (8+1)-photon ionization via the resonant 5$f$ and 4$f$ states, respectively. The blue curve is $[P_4(\cos\theta)]^2$. (b) The photoelectron angular distributions for the first-order ATI for intensities ranging from $1.3\times10^{13}$W/cm$^2$ to $2.0\times10^{13}$W/cm$^2$ (the intensity is the same as that in Fig. 2).

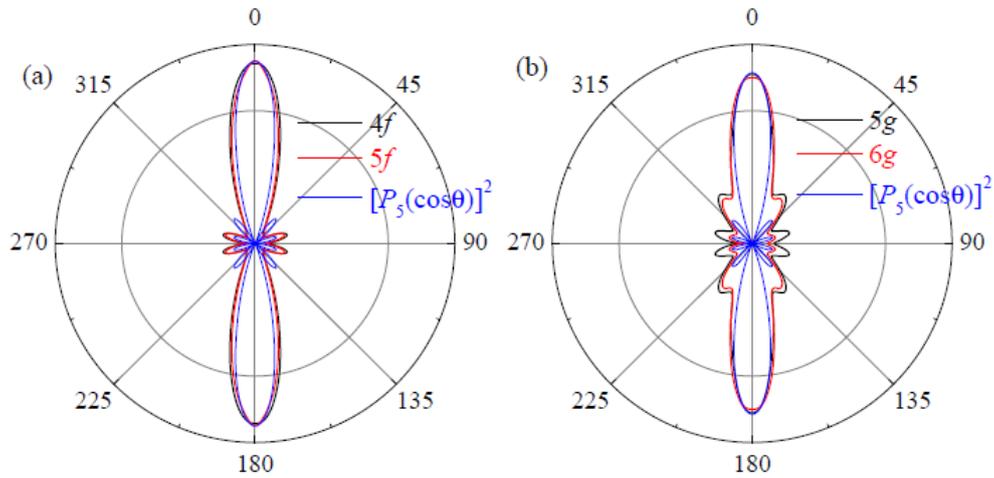

FIG. 5. (Color online) Polar plots of the photoelectron angular distributions of (a) (8+2)-photon ionization via the resonant 4$f$ and 5$f$ states and (b) (9+1)-photon ionization via the resonant 5$g$ and 6$g$ states. The blue curves are $[P_5(\cos\theta)]^2$.